%% file: main.tex
\documentclass[conference]{IEEEtran}
\input{common/packages}

\input{acronyms.tex}
\usepackage{array}
\newcolumntype{L}[1]{>{\raggedright\arraybackslash}p{#1}}
\usepackage{setspace}  
\begin{document}
\thispagestyle{empty}
\pagenumbering{gobble}
\singlespacing  
\title{Context-Aware Orchestration of Energy-Efficient Gossip Learning Schemes}
\vspace{-65pt}
\author{\IEEEauthorblockN{Mina Aghaei Dinani,\\Adrian Holzer}
\IEEEauthorblockA{University of Neuchatel,\\ Switzerland\\name.surname@unine.ch}
\and
\IEEEauthorblockN{Hung Nguyen}
\IEEEauthorblockA{The University of Adelaide,\\ Australia\\hung.nguyen@adelaide.edu.au}
\and
\IEEEauthorblockN{Marco Ajmone Marsan}
\IEEEauthorblockA{
Institute IMDEA Networks,\\ Spain\\ajmone@polito.it}
\and
\IEEEauthorblockN{Gianluca Rizzo}
\IEEEauthorblockA{HES-SO Valais, Switzerland, and\\
University of Foggia, Italy
\\gianluca.rizzo@hevs.ch}
\vspace{-25pt}
}
\maketitle
\thispagestyle{plain}
\pagestyle{plain}
\input{00_abstract}

\textbf{keywords --- Distributed Learning, Energy Efficiency, Gossip Learning, Opportunistic Communication}
\input{01_Intro}

\input{02_system_model}

\input{03-GL}
\input{04_problem_formulation}
\input{OGL_architecture}
\input{05-numerical_assessment}
\input{conclusion}

\end{document}

%% file: common/packages.tex
\usepackage{lineno,hyperref}
\usepackage{amsmath}
\usepackage{verbatimbox}
\usepackage{graphicx}
\usepackage{tabularx}
\usepackage[usenames, dvipsnames]{color}
\usepackage{rotating}
\usepackage{caption}
\usepackage{algorithmicx}

\usepackage{comment}
\usepackage{algorithm}
\usepackage[noend]{algpseudocode}
\makeatletter

\usepackage{amssymb}
\usepackage{subfig}
\newtheorem{theorem}{Theorem}

\let\labelindent\relax
\usepackage{enumitem}
\setlist[itemize]{align=parleft,left=0pt..1em}

\newtheorem{myproblem}{Problem}

\def\bd{\begin{definition}}
\def\ed{\end{definition}}
\def\bt{\begin{theorem}}
\def\et{\end{theorem}}
\def\be{\begin{center}\begin{equation}}
\def\ee{\end{equation}\end{center}}
\def\bc{\begin{corollary}}
\def\ec{\end{corollary}}
\def\bl{\begin{lemma}}
\def\el{\end{lemma}}
\def\br{\begin{remark}}
\def\er{\end{remark}}

\usepackage{multirow}

%% file: acronyms.tex
\usepackage{acronym}
\acrodef{cnn}[CNN]{Convolutional Neural Network}
\acrodef{nas}[NAS]{Neural Architecture Search}
\acrodef{cnn}[CNN]{Convolutional Neural Network}
\acrodef{iot}[IoT]{Internet of Things}
\acrodef{ml}[ML]{Machine Learning}
\acrodef{gl}[GL]{Gossip Learning}
\acrodef{fl}[FL]{Federated Learning}
\acrodef{tvg}[TVG]{Time-Varying Graphs}
\acrodef{dsrc}[DSRC]{Dedicated Short-Range Communications}
\acrodef{d2d}[D2D]{Device-to-Device}
\acrodef{v2v}[V2V]{Vehicle-to-Vehicle}
\acrodef{er}[ER]{Erdős-Rényi}

\acrodef{OGL}[OGL]{Optimized Gossip Learning }
\acrodef{DNN}[DNN]{Deep Neural Network }
\acrodef{tvg}[TVG]{time-varying graph}

%% file: 00_abstract.tex
\begin{abstract}
Fully distributed learning schemes such as Gossip Learning (GL) are gaining momentum due to their scalability and effectiveness even in dynamic settings. However, they often imply a high utilization of communication and computing resources, whose energy footprint may jeopardize the learning process, particularly on battery-operated IoT devices. 
To address this issue, we present \ac{OGL}, a distributed training approach based on the combination of GL with adaptive optimization of the learning process, which allows for achieving a target accuracy while minimizing the energy consumption of the learning process. 
We propose a data-driven approach to \ac{OGL} management that relies on optimizing in real-time for each node the number of training epochs and the choice of which model to exchange with neighbors based on patterns of node contacts, models' quality, and available resources at each node. Our approach employs a DNN model for dynamic tuning of the aforementioned parameters, trained by an infrastructure-based orchestrator function.
We performed our assessments on two different datasets, leveraging time-varying random graphs and a measurement-based dynamic urban scenario. Results suggest that our approach is highly efficient and effective in a broad spectrum of network scenarios.
\end{abstract}

%% file: 01_Intro.tex
\section{Introduction}
Distributed learning schemes are poised to become one of the key enablers of future 6G networks, as they allow fast and efficient training of complex and large-scale models while delivering better reliability and fault-tolerance than traditional, centralized approaches. 
Among these, Gossip Learning (GL) schemes are of special interest, as they do not require uploading models to a parameter server, thus offering better robustness and scalability.

Originally introduced in \cite{ormandi2013gossip}, GL schemes train ML models over decentralized data via direct model gossiping among nodes. Several versions of GL have been proposed for dynamic settings \cite{10266757,dinani2021gossip}. In these works, the changing topology is a result of varying patterns of network connectivity, changes in node availability (e.g. due to node duty cycling or battery depletion), and churn, among others, as it is often the case in realistic mobile edge and vehicular scenarios and use cases. GL is based on a combination of iterative local training, and model exchange over wireless channels. Both tasks on battery-operated, resource-constrained IoT and edge devices might imply rapid energy budget depletion, potentially slowing down and jeopardizing the whole learning process.

Recently, several works have focused on decreasing the energy footprint of distributed learning, albeit in server-based architectures such as Federated Learning (FL).
~\cite{zhou2021communicationefficient} reduces the amount of exchanged models in FL by decreasing the number of communication rounds. Other approaches focus instead on reducing the number of exchanged models at each round. These methods consider factors such as the size and quality of a node’s local dataset~\cite{Abdelmoniem_2023, marfoq2022personalized, malandrino2021federated,wu2021fastconvergent}, the rate of network evolution~\cite{zhou2021communicationefficient, wang2019adaptive}, and the node's trustworthiness~\cite{imteaj2021fedar}.
All these works, however, consider a static network. 
~\cite{dinani2022vehicle} proposes a Gossip Learning scheme, evaluating the effect of the number of models merged at each round in a vehicular network. It shows that the resource-optimal value for these parameters is highly context-specific. However, it considers measurement-based mobility patterns, making it hard to untangle dependencies between mobility features and the performance of the training scheme. All these works leave unanswered the critical question of when and which nodes should exchange models in a fully distributed learning scheme to achieve a target performance (in terms of accuracy and convergence speed) in a dynamic network in an energy-optimal manner.

In this paper, we consider a scenario of reference in which cellular connectivity is pervasive, and it allows taking advantage of an orchestration function that monitors the gossip-based learning process without requiring infrastructure-based exchanges of training data or ML models. 
The primary contributions of this paper are: 
\begin{itemize}[topsep=0pt]
    \item  We propose \ac{OGL}, a gossip-based training strategy for dynamic networks capable of adapting to a wide range of network topologies and dynamic settings.
    \item We present a data-driven approach for the dynamic management of \ac{OGL}, which achieves a target performance in a resource-efficient manner by proactively adapting the models' distribution and training parameters to local conditions. The approach employs a \ac{DNN} model, which is trained in an offline manner and distributed to all nodes at the start of the learning process, and that enables each node to tune the main parameters of the OGL scheme adaptively. 
   \item We assess the effectiveness and efficiency of our approach by leveraging time-varying random graphs and a measurement-based mobility trace. These results suggest that it substantially outperforms a set of baseline approaches while achieving the target minimum accuracy in all of the considered scenarios. 
\end{itemize}

%% file: 02_system_model.tex
\section{System model}
\label{sec:System model}
We consider a set $V$ of nodes, with cardinality $|V|$, (modeling e.g., mobile devices, UAVs, or connected vehicles) moving within a specific region according to an arbitrary mobility model and during a predefined time interval $T$ (the \textit{observation interval}). Let $v \in \mathbb{N}$ denote the unique identifier of a node. We assume nodes can communicate directly with each other through wireless peer-to-peer (P2P) communications (e.g., using  DSRC or Bluetooth Low Energy \cite{gomez2012overview}). Furthermore, each node is equipped with a cellular network interface. Two nodes can exchange information whenever they are in \textit{contact}, that is, within each other's transmission range. We assume these exchanges are always unicast (one-to-one). However, the proposed scheme can be easily extended to incorporate the effects of multicasting and broadcasting. We assume there is a coordination function (possibly implemented by a Software Defined Network Controller(SDNC)~\cite{sdn}) in the region, and it resides within the cellular access network. The coordination function comprises an auxiliary ML model $M_{tune}$. The coordination function, through its cellular network interface, transmits $M_{tune}$ to the nodes entering the region. Hence, the architecture of this model is equal for all nodes. Note that all communications are P2P, except for the initial dissemination of  $M_{tune}$ model from the coordination function to nodes. Every node independently employs $M_{tune}$  to fine-tune and adjust its learning parameters. Moreover, we assume each node entering the region possesses a local model $w_v$  and a  \textit{local dataset}, partitioned in the training set  $\mathcal{D}_{v}$, and validation set  $\mathcal{S}_{v}$, which generally differ in size and composition for each node.  The choice of the validation and training set size is context-specific. 
We also assume that the observation interval is segmented into $I$ \textit{slots} of equal size, short enough that node mobility patterns can be considered not to vary substantially within each slot. Let $t\in 1,..,I$ be the label of slots.

%% file: 03-GL.tex
\section{The OGL Approach to Energy-Efficient Gossip Learning }
\subsection{A Gossip-Based Collaborative Training Algorithm} \label{sec:Deep_GL}
We assume all nodes in the region train their local model through a gossip-based cooperative learning algorithm, denoted as \ac{OGL}, and based on P2P model exchanges among nodes.  Such an approach is orchestrated by a cellular-based coordination function, without requiring the exchange of the trained models or training data (which would potentially expose it to privacy breaches) between each node and the coordinator. 
At the beginning of the scheme, all nodes present in the region randomly generate an initial
local model $w_0$. We assume the generation procedure to produce the same random initial model for all nodes. Similarly, after the beginning of the scheme, whenever a node enters the region, it generates $w_0$ following the same procedure.  Starting from $w_0$, every node elaborates a ML model (which we assume will be used by the node itself. e.g. to carry out the same inference task, e.g. trajectory prediction, or image recognition) by alternating local training on each node's local training set, with model aggregation with models received from neighbors. Moreover, to all nodes joining the scheme, the coordination function delivers an \textit{auxiliary ML model} $M_{tune}$.  
 Each node employs $M_{tune}$ for the adaptive tuning of some key parameters of the learning process. Such dynamic management of the learning process at each node is based on each node's available hardware resources and power budget. In addition, it also accounts for each node's context in terms of the number of neighbours, the speed at which they vary over time, and the quality and quantity of their local model (i.e., in terms of mean accuracy or loss), among others. Each node then uses the $M_{tune}$ to modulate the number of local training epochs and to choose, among its neighbours, those whose local model should be requested and used for improving the node's local model, as we will explain later.

Then, at every time slot, the \ac{OGL} algorithm proceeds through three \textit{phases}.
The duration of each phase can be tuned and adapted to the specific training task and setup, and it does not need to be synchronized across nodes.

In the \textit{training} phase, each node in the region applies $M_{tune}$ to get the number of epochs $Z_{v,t}$  that it has to train its local model over its local dataset and train its model accordingly.  Subsequently, it assesses its local model over its validation set $\mathcal{S}_{v}$ to derive the loss value $l_v$ used in the next phase. The choice of the loss function is context-specific.

In the \textit{communication} phase, nodes exchange the loss value of their local model with their neighbours. Each node then employs $M_{tune}$ to identify the neighbouring nodes from which it should request the transfer of their local models. The node then initiates a request directed towards the selected nodes, soliciting their respective models. In response, these requested nodes transmit their models if they are still within range of each other. During this phase, a node may not request any model, e.g. because it has no neighbours or when the $M_{tune}$ indicates that no model is worth requesting among the available neighbours' models. We assume that the connectivity between nodes is relatively stable while exchanging models. 

Finally, in the \textit{merging} phase, each node combines the models received from the chosen neighbours and its local model to produce a new version of its local model. The merging procedure consists of a weighted averaging method. The weights associated with each merged model are computed via the DFed Pow strategy~\cite{dinani2021gossip}. In DFed Pow, the weight of each model to be merged is a function of the inverse of loss calculated on the node's validation set. Note, however, that our approach is more general and does not rely on a specific algorithm for calculating the weights for merging.

The three phases are repeated at each time slot until a stopping condition is met (e.g., after a maximum number of iterations, when the average local models' accuracy surpasses a certain threshold or when there is no significant improvement in the model’s accuracy over several rounds). Regardless of the reason, the final round at which the algorithm stops is called the cut-off round. \nolinebreak

%% file: 04_problem_formulation.tex
\subsection{Formulation of the energy optimization problem}
The main goal of our OGL approach is to enable the energy-efficient training of an ML model in a distributed manner. As mentioned, this is enabled by an orchestrator function that elaborates and distributes the $M_{tune}$ model among all the nodes entering the region.  Given the node context, as well as some key parameters of the system and the training task, such a model enables each node to tune the number of training epochs and the set of ML models to merge, which allows for achieving a given target accuracy while minimizing a cost function which models the overall energy cost of the training process.

In what follows, we formalize the energy optimization problem that the orchestration function tries to solve. The cost function we consider is the sum of two components. The first one accounts for the computing costs. Generally, the energy consumption associated with a computation task is determined by CPU(or GPU) usage and memory resources \cite{9045988}. Those, in turn, depend upon the architecture implemented, the quantity of data it processes, and the node's characteristics. In this work, we assume all nodes have the same computing power. Then, the energy required to run the (local) training process is:%
\begin{equation}
\label{'eq:local_training'}
   S(Z)= \sum_{t \in T}\sum_{v \in V} Z_{v,t} d_v (e_g+ e_s)
\end{equation}%
\begin{itemize}
\item $Z_{v,t}$ depicts the number of epochs required by node $v$ to train its local model using the local dataset at time slot $t$;
\item $e_g$ is the energy consumed by the CPU or GPU to perform one  training epoch on one sample;
\item $e_s$ is the energy required to provide storage and memory resources for the training of one epoch on a sample;
\item $d_v$ denotes the number of samples of the local training set of node $v$;
\item $T$ is the label of the slot at which convergence happens.
\end{itemize}%
The energy consumed for computing the loss of the local model on the validation set is modelled as follows:%
\begin{equation}
   \Gamma = \sum_{t \in T}\sum_{v \in V} s_v (e_e+ e_{es})
\end{equation}%
$e_e$ and $e_{es}$ are the energy consumed by the CPU (or GPU) and energy required to provide storage and memory to evaluate the local model on one dataset sample. $s_v$ indicates the size of the validation set at node $v$.

The second component of the cost function accounts for the communication costs. The communication costs consider the exchanges between nodes:%
\begin{equation}
\label{eq:DGL_communication}
   C(k) = C^{d2d}\sum_{t \in T}\sum_{v \in V}  h_{v,t} L +  k_{v,t} (M+R)
\end{equation}%
\begin{itemize}
\item $C^{d2d}$ is the cost per byte of a d2d (peer to peer) transfer  
\item $\mathcal{H}_{v,t}$ is the set of neighbors of node $v$ at time slot $t$, of cardinality $h_{v,t}$; 
\item $\mathcal{K}_{v,t}\subseteq H_{v,t}$ is the set of chosen neighbours of node $v$ at time slot $t$ from which models to be merged are retrieved, of cardinality $k_{v,t}$.
\item $L$, $R$ and $M$ are the message size containing loss value, request and a local model, respectively.
\end{itemize}%
Note that such a cost function neglects the cost of model merging, as it is usually negligible \cite{dinani2021gossip,dinani2022vehicle}. Let $\mathcal{Z}=\{Z_{v,t}\}$, and $\mathcal{K}=\{\mathcal{K}_{v,t}\}$. Thus, an optimal \ac{OGL} orchestration scheme is a solution to the following optimization problem:%
\begin{myproblem}
\label{prob:1}
\begin{equation}\label{eq:objectivefunction}
\underset{\mathcal{Z},\mathcal{K}}{\text{minimize}}\ C(k) + \beta (S(Z)+\Gamma)
\end{equation}
\quad Subject to:
	\begin{align}
      r&\geq r^{0}
	\end{align}
\end{myproblem}%
Where $r$ denotes the mean accuracy of the trained model across all nodes achieved at convergence, and $r^{0}$ is its target minimum value. By varying $\beta$, it is possible to adapt the cost function to settings with different resource availability on user devices and at the cellular network and to different incentive schemes for resource sharing and cooperation.\nolinebreak

%% file: OGL_architecture.tex
\subsection{OGL architecture, components and functions}
\label{sec:aux_ml}
The OGL approach aims at solving Problem \ref{prob:1} by training a \ac{DNN}-based auxiliary ML model, which enables nodes to adapt in real-time the learning process to available contributions by neighbours and, more generally, to each node's context. The auxiliary model is trained by the orchestrator, on a training dataset whose data points are labelled by simulating the system.

We assume the orchestrator regularly collects data from every node, to be used as the feature set of the auxiliary model that it has to train. The data collected are those that are well known from the state of the art to be relevant to the training process and its efficiency. These include computing and communication costs, the number of neighbours for each node at each time slot, the size of the local dataset, the available computing power, and the initial power budget of each node. Such a choice of features as input parameters for the auxiliary model is, however, one of many possible, and our approach is independent of it.  From each set of input parameters, the orchestrator derives a set of full system configurations by associating to the input parameters a random value for each of the parameters in $\mathcal{Z}$ and a random subset $\mathcal{K}$ of each node's neighbours.  These inputs are fed to a simulator, which labels them with the outputs and performance metrics of the distributed training scheme. Specifically, for every node and every time slot in a given time interval during which the orchestrator has collected data, the simulation derives the local model accuracy, the loss value, and the energy budget of each node. In such a way, a training set is produced, which is then used to train a \ac{DNN} model.

This model is trained and evaluated using a k-fold cross-validation approach ~\cite{Refaeilzadeh2009}, with  $10$ folds. The architecture of the model is a multi-layer perceptron composed of four layers. The multiple layers allow models to be more efficient at learning complex features ~\cite{Kruse2013}. The initial layer is a dense layer with 64 neurons and a rectified linear unit (ReLU) activation function. The second layer is a flattening layer, which reshapes the input to a one-dimensional array. The third and fourth layers are dense layers with 32 and 16 neurons, respectively, and ReLU activation functions. The final layer is a dense layer with two neurons. The model is compiled with a mean squared error loss function and the Adam optimizer.  To prevent overfitting and save the best model, early stopping and model checkpoint callbacks are used. 
This approach ensures a robust evaluation of the model performance, as it assesses the model’s ability to generalize to unseen data.
Note that the selection of parameters in the model is either empirical or based on extensive usage in the state-of-the-art models. After the training and evaluation process, the best-performing model, referred to as $M_{tune}$, is saved for future use. This model encapsulates the optimal parameters learned during the training process. At runtime, the orchestrator function disseminates the  $M_{tune}$ model to all nodes entering the region. Then, at each time slot, each node feeds the $M_{tune}$ model with its own data, to determine the optimal number of local training iterations (number of epochs) and the optimal set of models to merge to achieve the given target accuracy while minimizing the energy cost of the whole process.

%% file: 05-numerical_assessment.tex
\section{Numerical assessment}
\label{sec:Performance evaluation}
\begin{table}[t!]
\small
 \begin{center}
\begin{tabular} {|L{3cm}|L{2.5cm}|L{2.5cm}|} 
\hline
\rule{0pt}{10pt}CNN parametrs& MNIST & CIFAR-10  \\\hline
\rule{0pt}{8pt}Input shape &(28,28,1) &(32,32,3) \\
\rule{0pt}{8pt}Batch size &32 & 64 \\
\rule{0pt}{8pt}Learning rate &0.0001 &0.001 \\
\rule{0pt}{8pt}Number of neurons &100 & 100 \\
\rule{0pt}{8pt}Momentum & 0.9&0.60 \\
\rule{0pt}{8pt}Kernel dimension & 3&3 \\
\rule{0pt}{8pt}Number of filters & 32 & 32 \\
\rule{0pt}{8pt}Number of outputs & 10&10 \\
\hline
\end{tabular}
\end{center}
\caption{Parameter values used to train the CNN model on the  CIFAR-10 and MNIST datasets.}
\label{tab:CNN_parametrs}
\end{table}
 \begin{figure}[t!]
 \centering
 \includegraphics[width=\columnwidth]{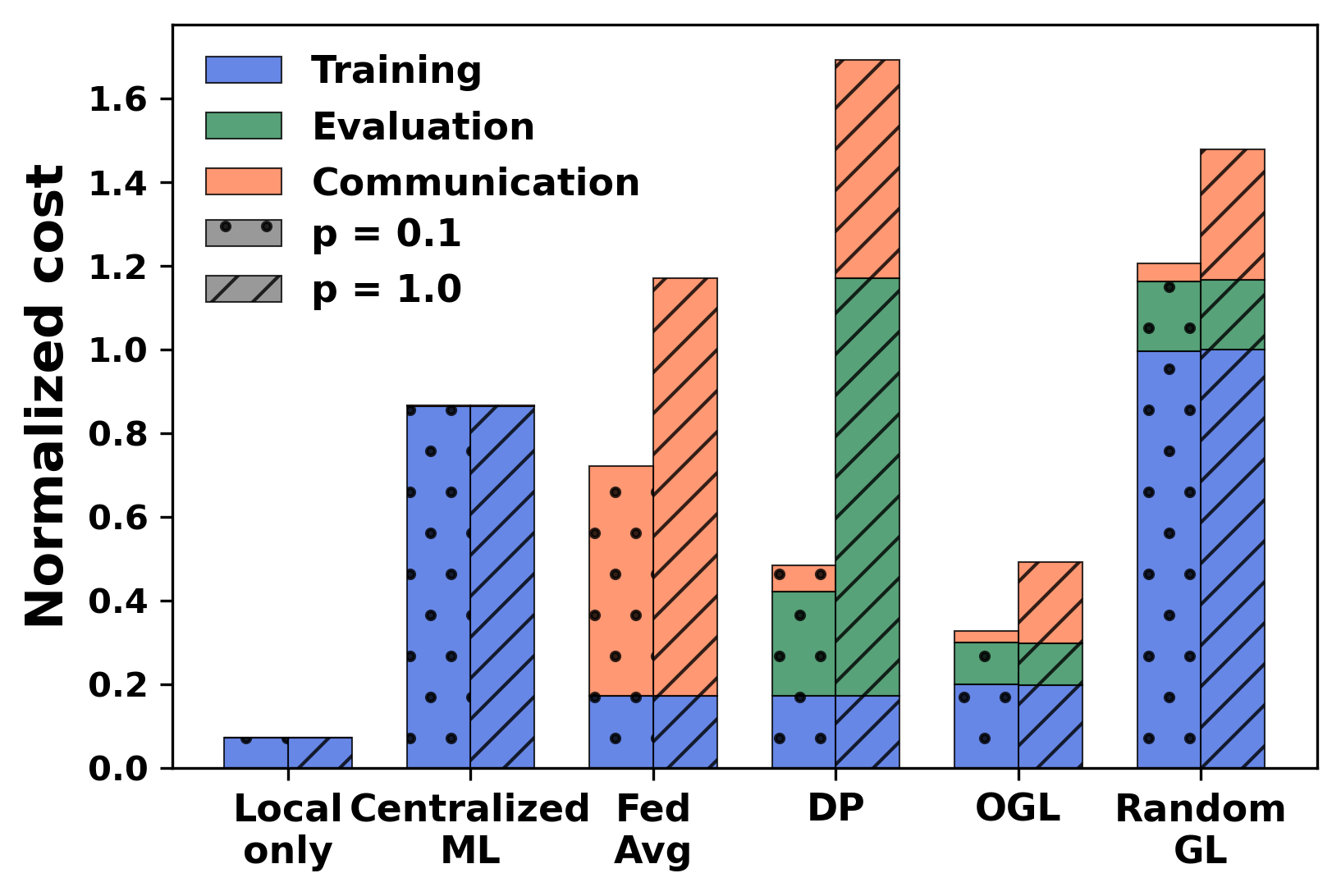}
 \caption{\small   Comparative analysis of cost function on the MNIST dataset with $|V|=6$ and various values for $p$. \normalsize}
 \label{fig:cost_comparisons}
 \end{figure}
\begin{table}[t!]
 \begin{center}
 \small
\begin{tabular}  {|L{2.2cm}|L{0.35cm}L{0.35cm}L{0.45cm}L{0.8cm}L{0.5cm}m{1cm}|}
\hline
\rule{0pt}{12pt}Algorithm &Acc &F1  &Loss &Precision&Recall &Cut-off round \\\hline
\rule{0pt}{8pt}Centralized ML &0.87 &0.87 &  0.45 & 0.88 & 0.87& 300\\
\rule{0pt}{8pt}Fed Avg &0.85 &0.84 & 0.57 & 0.87 & 0.85 &500\\
\rule{0pt}{8pt}Local only &0.36& 0.42 &  3.2 & 0.42 & 0.56 &200\\
\rule{0pt}{8pt}DP & 0.58&0.60 & 1.42 & 0.68 & 0.67&500\\
\rule{0pt}{8pt}\ac{OGL} &\textbf{0.88} &\textbf{0.88} &  \textbf{0.44}& \textbf{0.88} & \textbf{0.89}&320\\
\rule{0pt}{8pt}Random GL & 0.51&0.56  & 1.73 & 0.7& 0.6 & 500\\
\hline
\end{tabular}
\end{center}
\caption{Performance metrics of \ac{OGL} at the cut-off round, compared to baselines on the MNIST dataset, with  $|V|=6$ and $p = 1$. Results are presented with a $98\%$ confidence interval and a maximum margin of error of  $1\%$.}
\label{tab:mnist_metrics}
\end{table}
 \begin{figure}[t!]
 \centering
 \includegraphics[width= 9 cm , height = 5 cm]{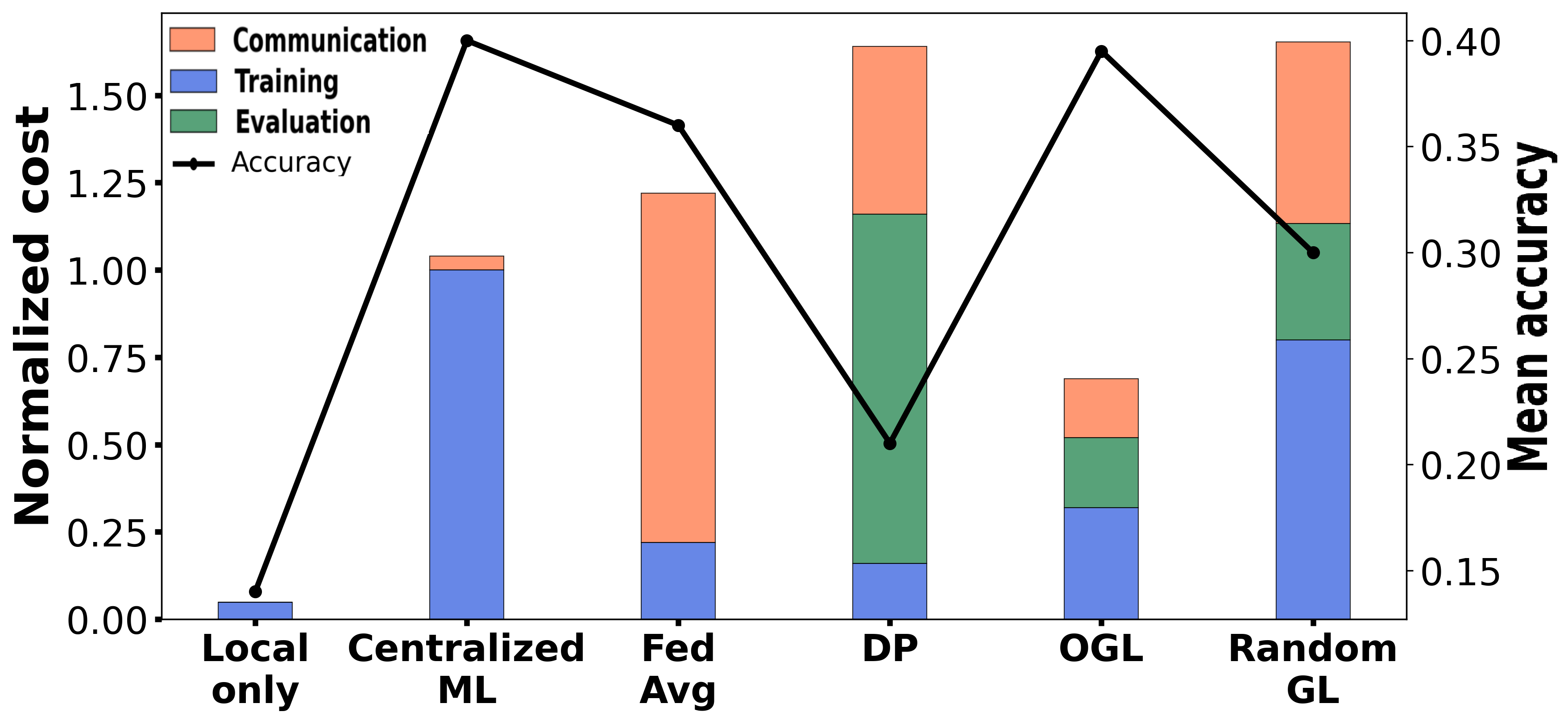}
 \caption{\small   Comparative analysis of cost function and mean accuracy at convergence on the CIFAR-10 dataset, with $|V|=6$ and $p=1$. \normalsize}
 \label{fig:cost_acc_cifar}
 \end{figure}
 \begin{figure}[t!]
 \centering
 \includegraphics[width=\columnwidth]{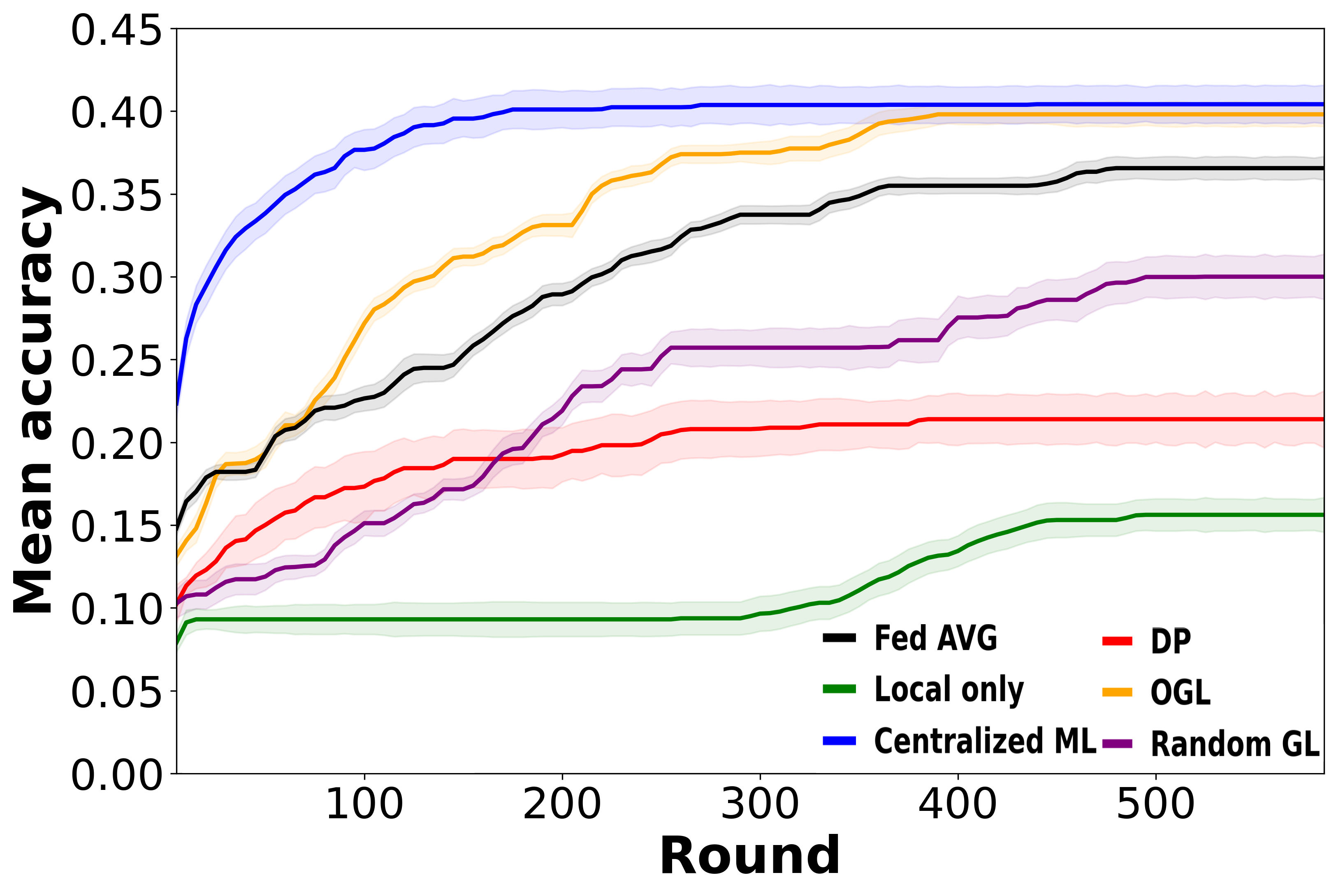}
 \caption{\small  Mean accuracy versus round of our OGL algorithm compared to baselines using the CIFAR-10 dataset, with $|V|=6$ and $p = 1$. Curves are associated with a $95\%$ confidence interval. \normalsize}
 \label{fig:comparative_acc_cifar}
 \end{figure}
\begin{figure*}[ht!]
     \centering
     \subfloat[\centering $|V| = 3, p=0.1 $]{\includegraphics[width=.26\textwidth]{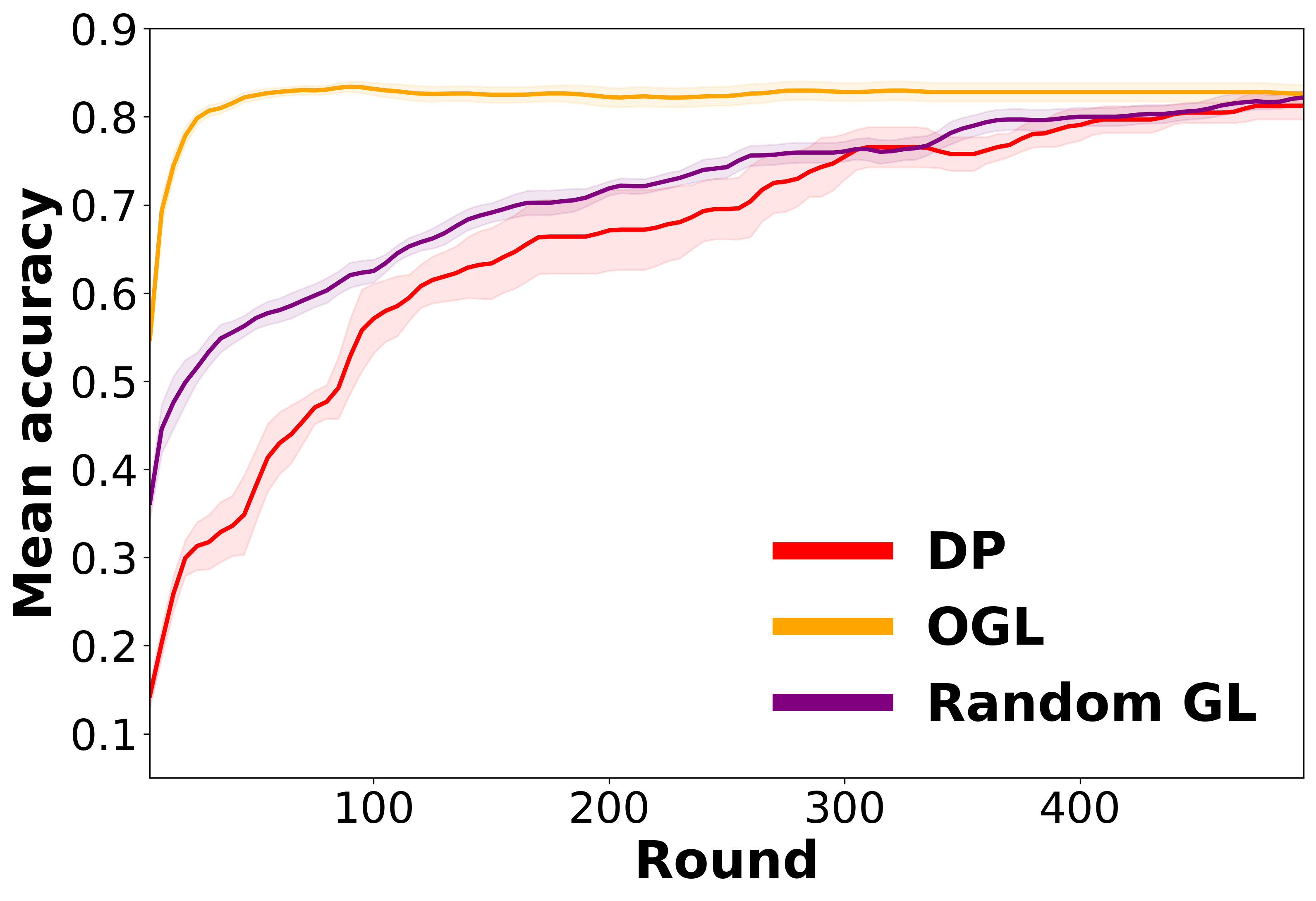}}
      \subfloat[\centering $|V| = 3, p= 0.7$]{ \includegraphics[width=.25\textwidth]{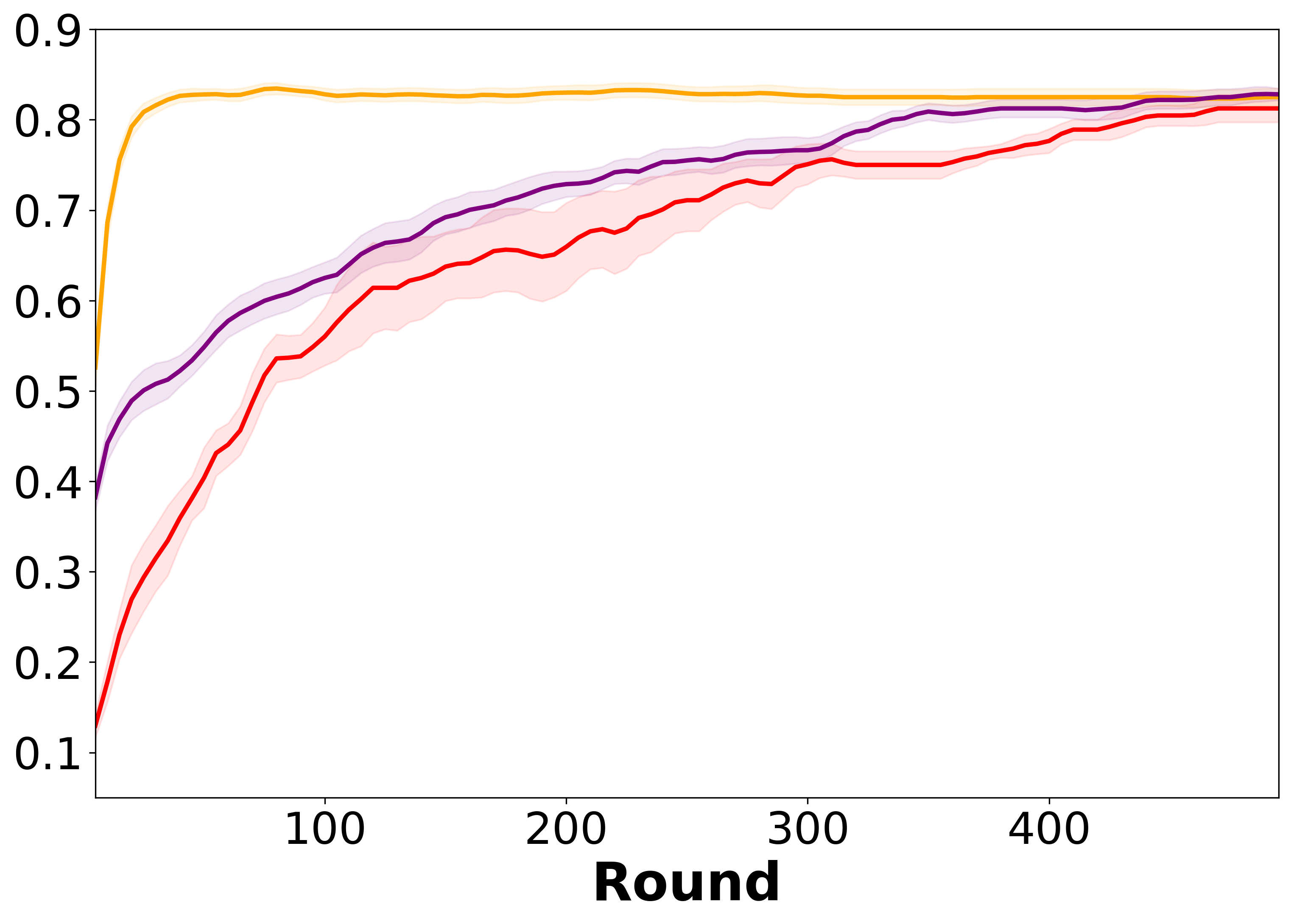}}
        \subfloat[\centering $|V| = 12, p=0.1 $]{\includegraphics[width=.25\textwidth]{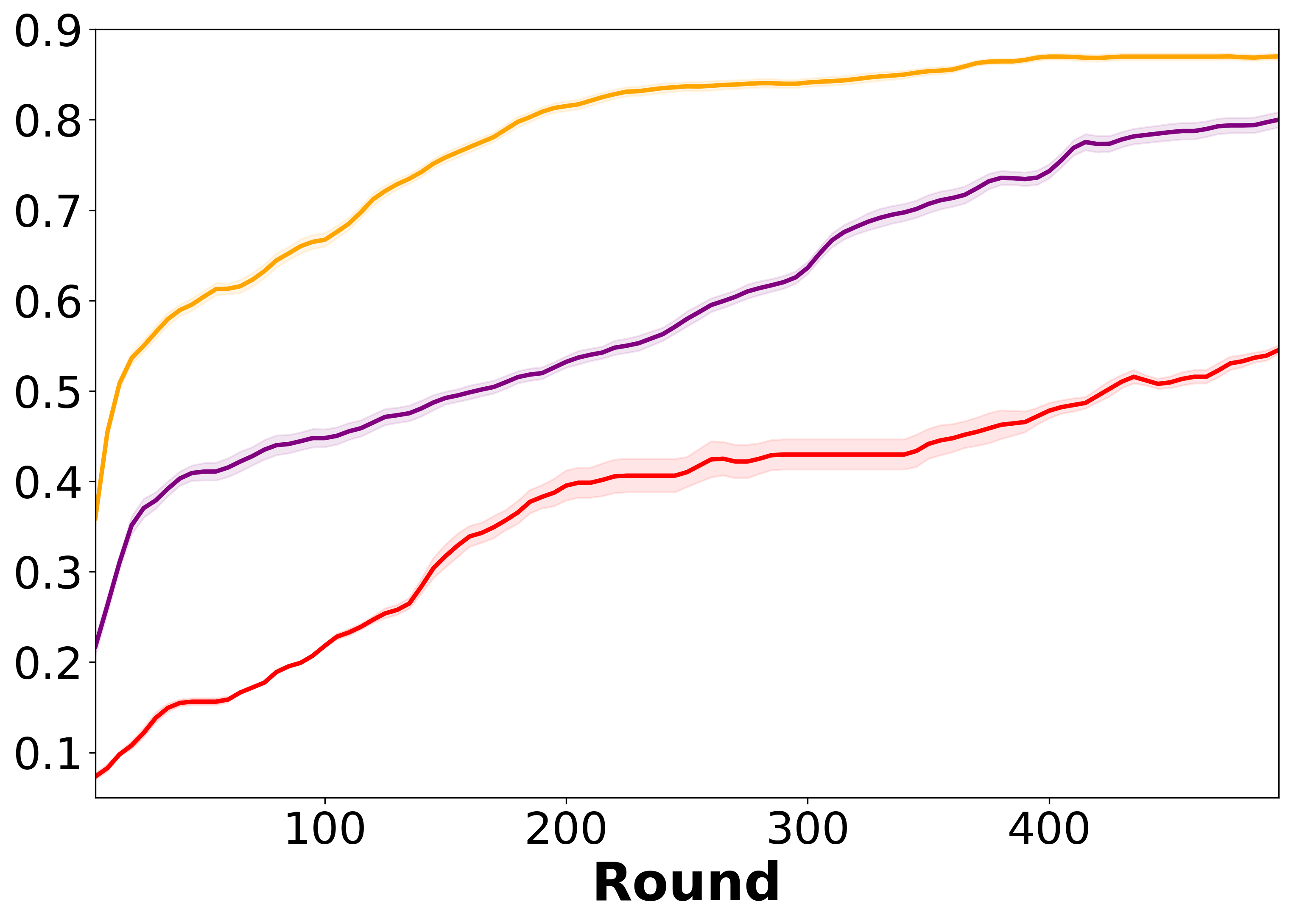}}
        \subfloat[\centering $|V| = 12, p=0.7$]{ \includegraphics[width=.25\textwidth]{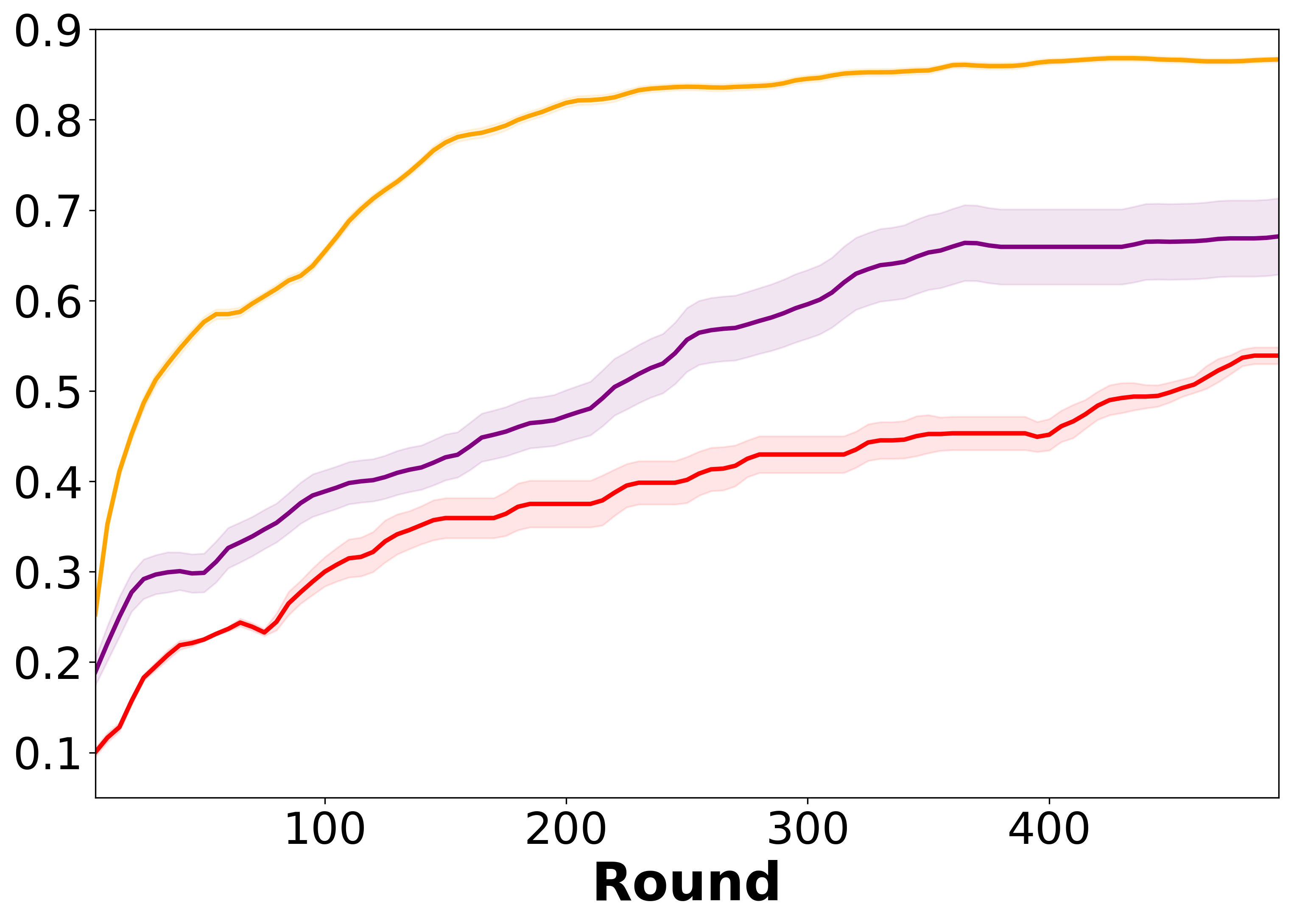}}
      \caption{Mean accuracy versus round for \ac{OGL}, random GL, and DP algorithm using the MNIST dataset, for different values of number of nodes in the system and edge probability. Plots are with $95\%$ confidence interval.}  
      \label{fig:mnist_comparisons}
 \end{figure*}
 \begin{figure*}[ht!]
     \centering
     \subfloat[\centering $|V| = 3, p=0.1 $]{ \includegraphics[width=.26\textwidth]{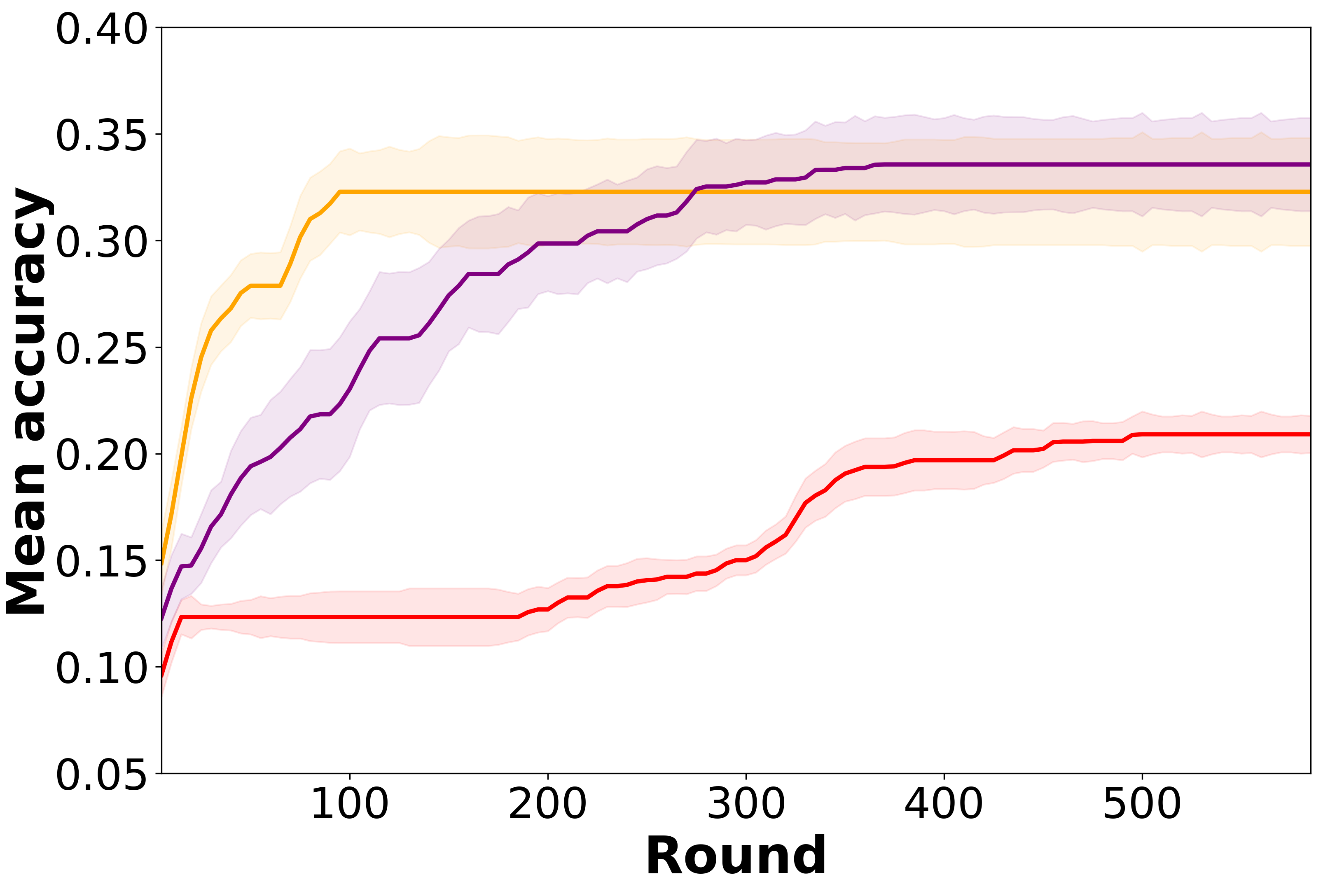}}
     \subfloat[\centering $|V| = 3, p=0.7$]{\includegraphics[width=.25\textwidth]{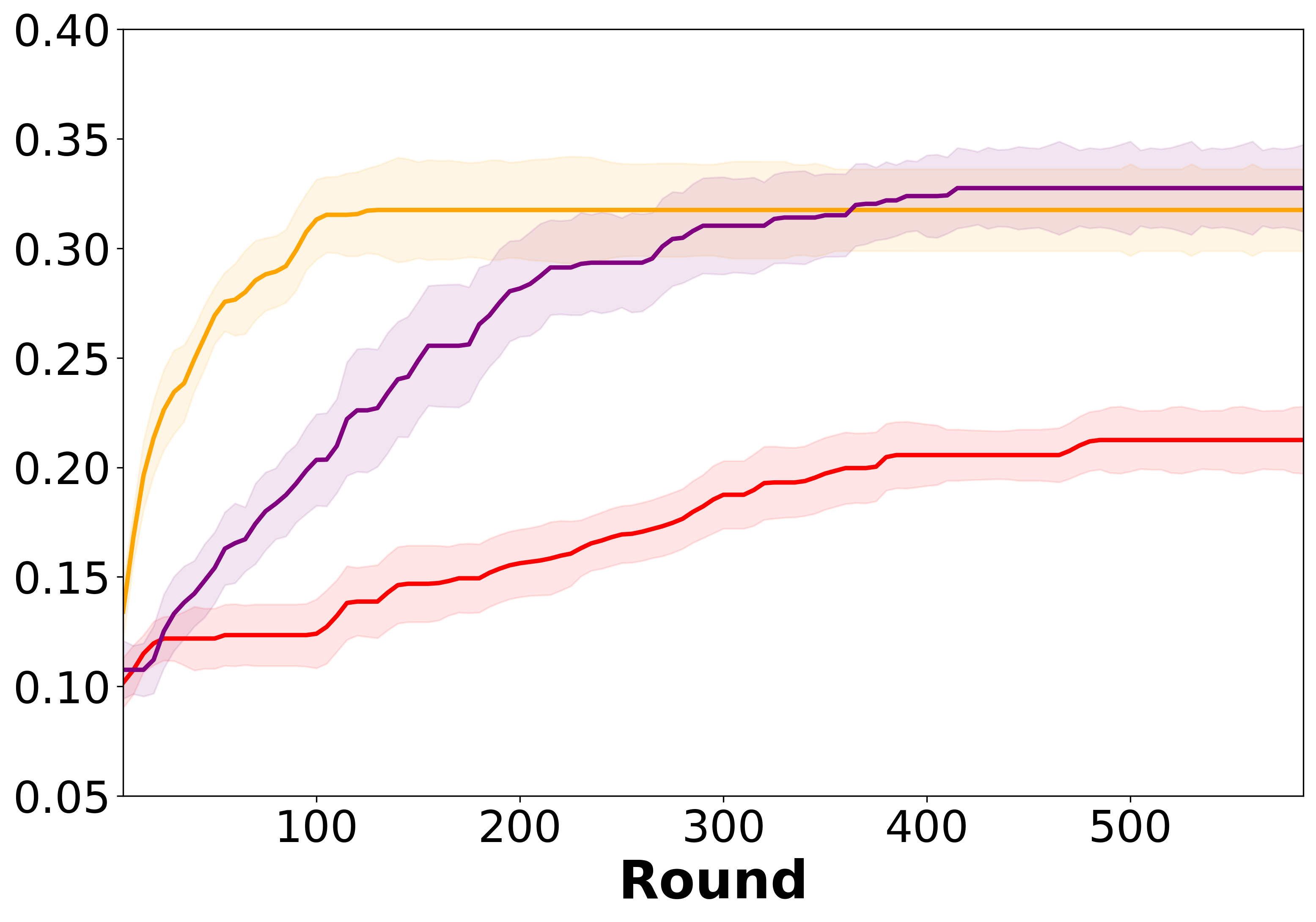}}
     \subfloat[\centering $|V| = 12, p=0.1 $]{\includegraphics[width=.25\textwidth]{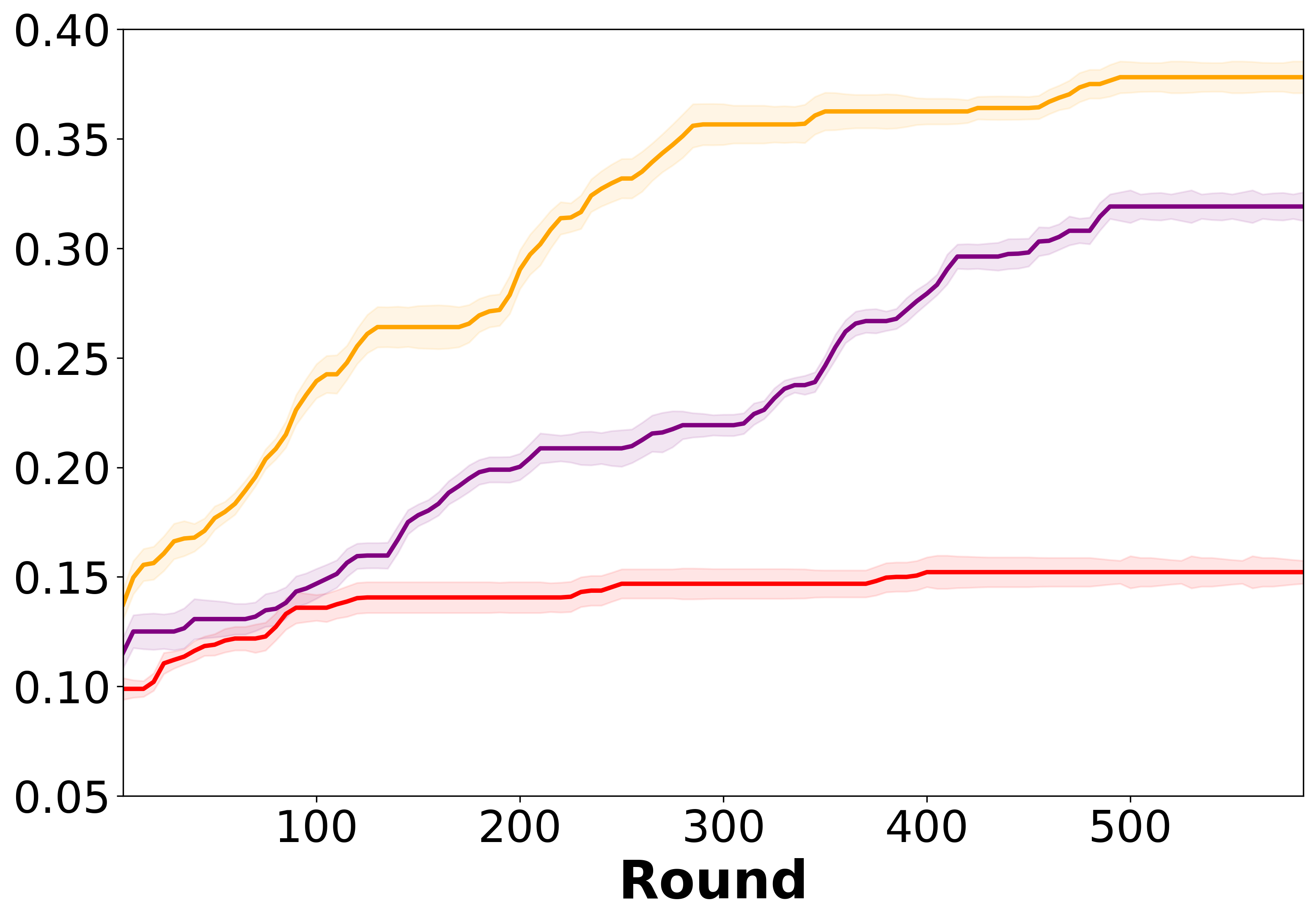}}
     \subfloat[\centering $|V| = 12, p=0.7$]{\includegraphics[width=.25\textwidth]{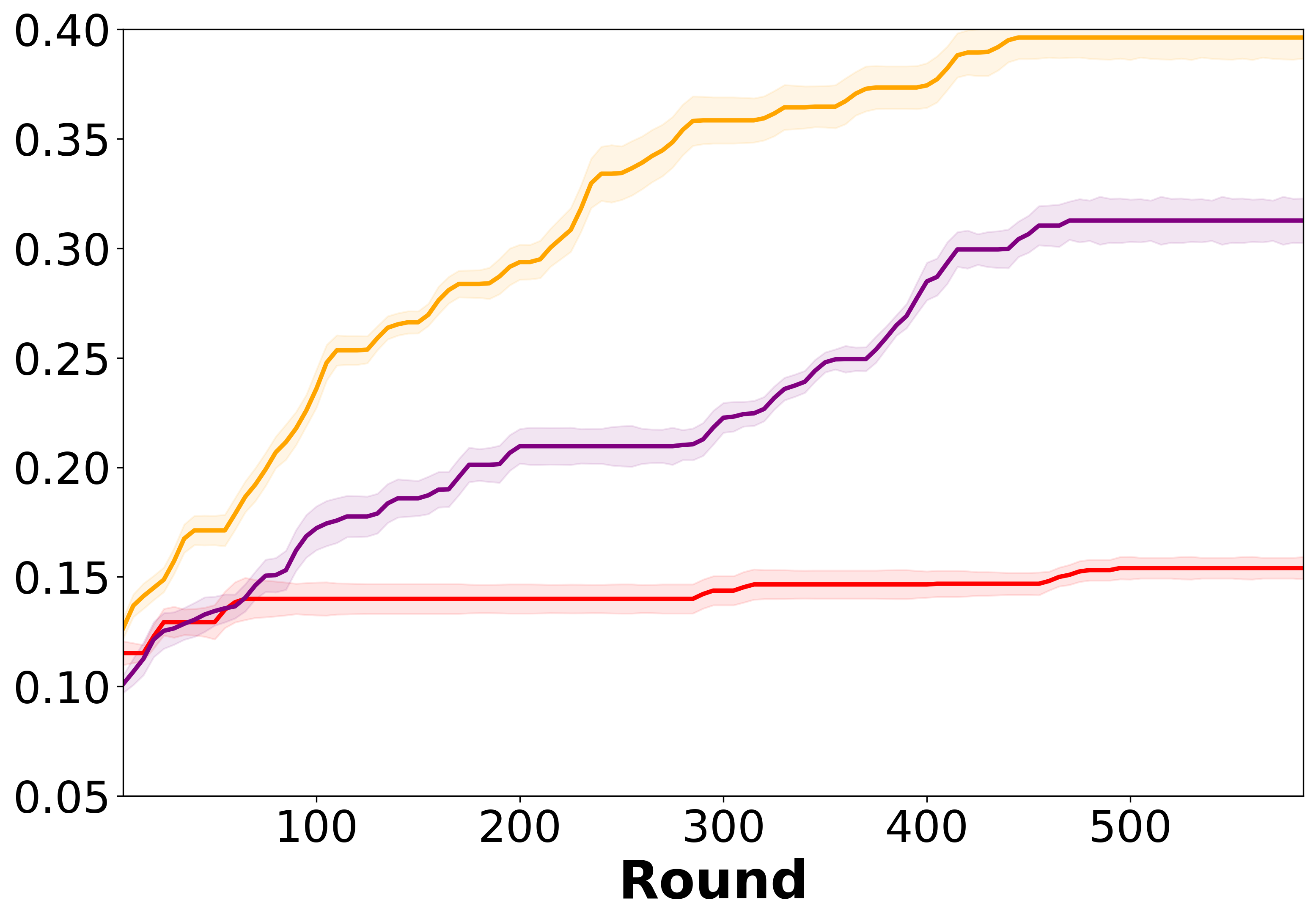}}
     \caption{Mean accuracy versus round for the \ac{OGL}, Random GL, and DP algorithms, for the  CIFAR-10  dataset, for different values of number of nodes in the system and edge probability. Plots are with $95\%$ confidence interval.}  
     \label{fig:cifar_comparisons}
\end{figure*}
  \begin{figure}[t!]
 \centering
 \includegraphics[width=\columnwidth]{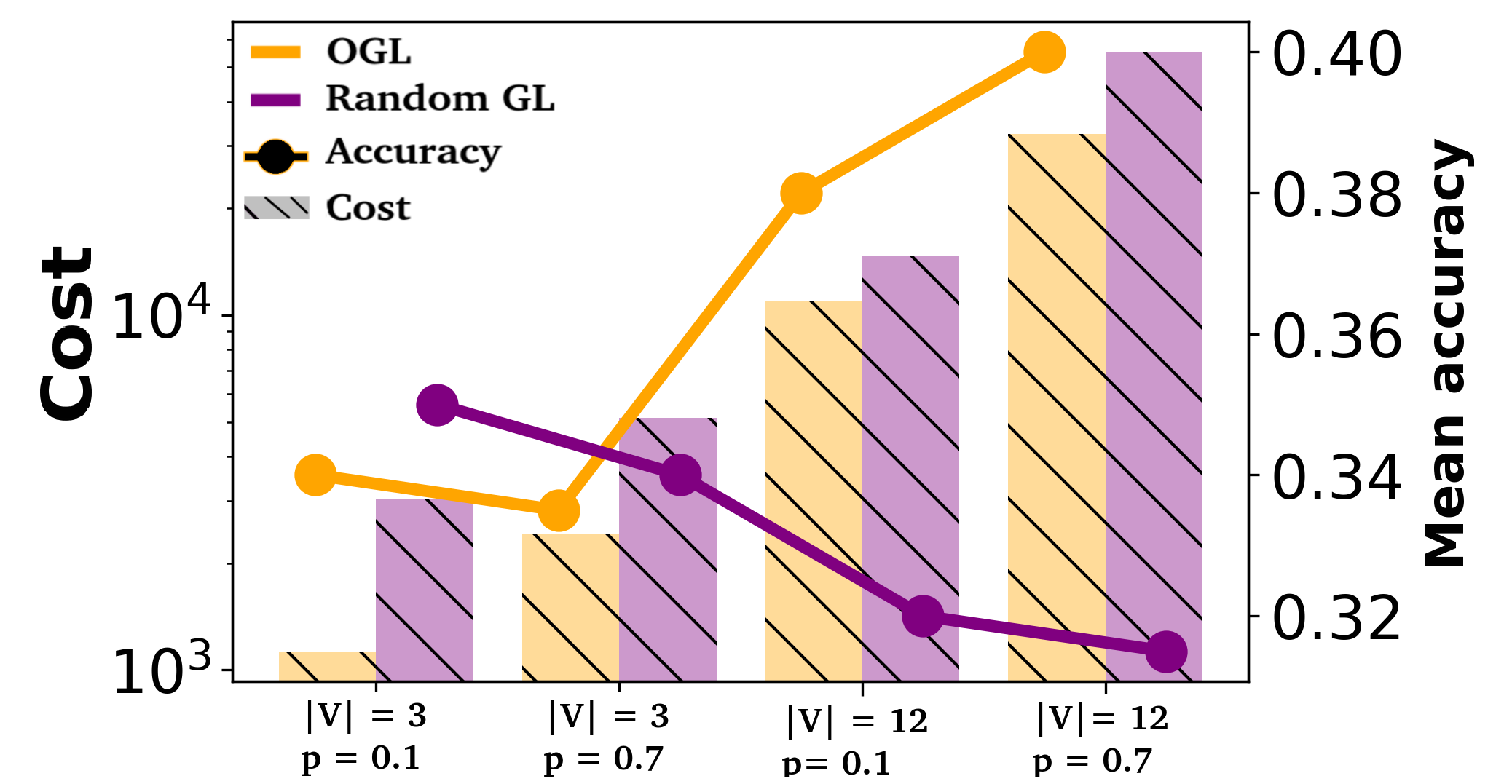}
 \caption{\small Comparison of mean accuracy at convergence time versus cost for the OGL and random GL algorithms over different network configurations using the CIFAR-10 dataset. Results are presented with a $98\%$ confidence interval and a maximum margin of error of  $2\%$.\normalsize}
 \label{fig:bar_cost_acc_cifar}
 \end{figure}
  \begin{figure}[t!]
 \centering
 \includegraphics[width=8 cm , height = 5 cm]{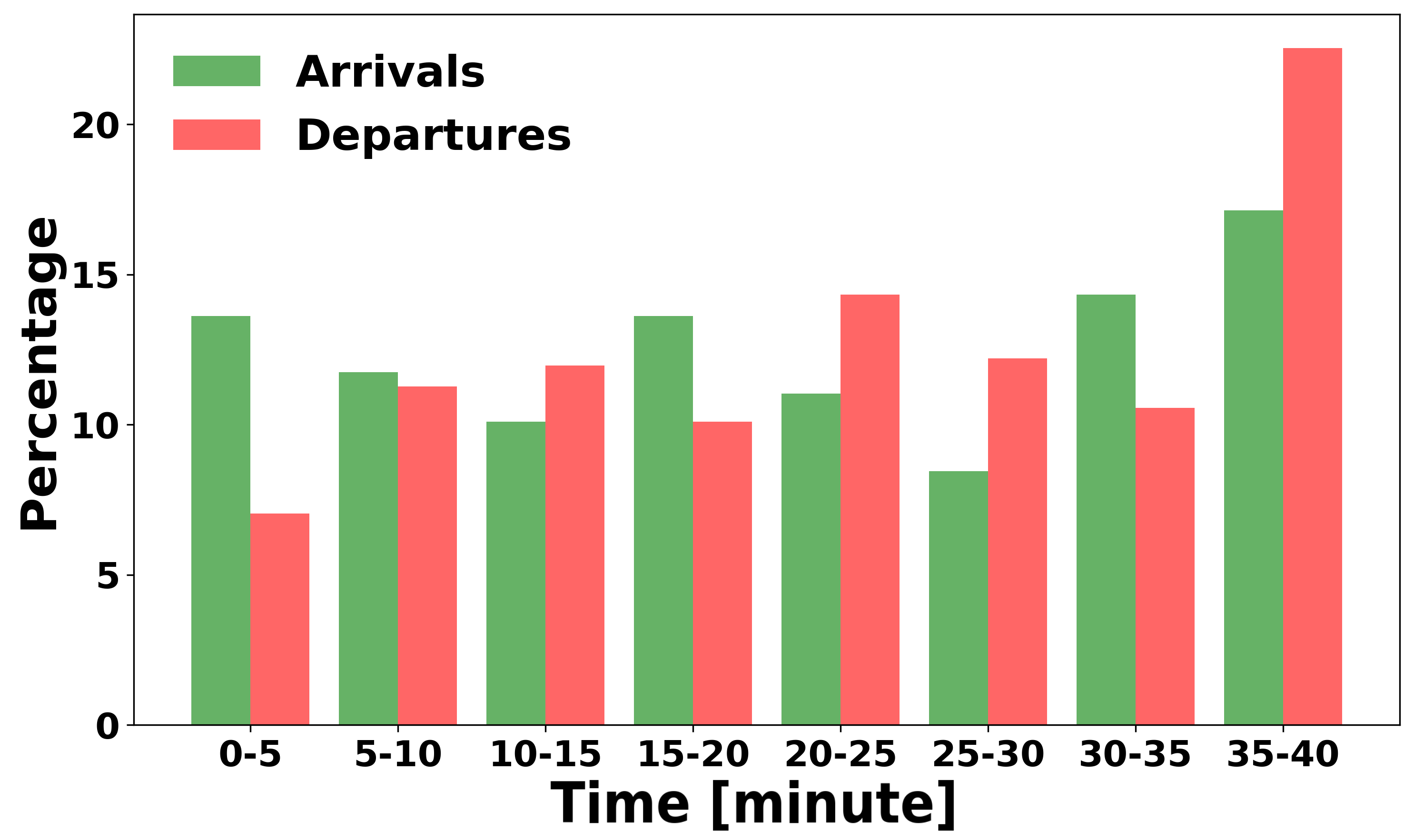}
 \caption{\small Histogram representing the percentage of vehicle arrivals, and departures across different time slots in the Luxembourg City off-peak scenario (12:00-12:40 PM). Time on the x-axis shows the time from the beginning of the scheme.\normalsize}
 \label{fig:histogram_churn}
 \end{figure}
  \begin{figure}[t!]
 \centering
 \includegraphics[width=\columnwidth]{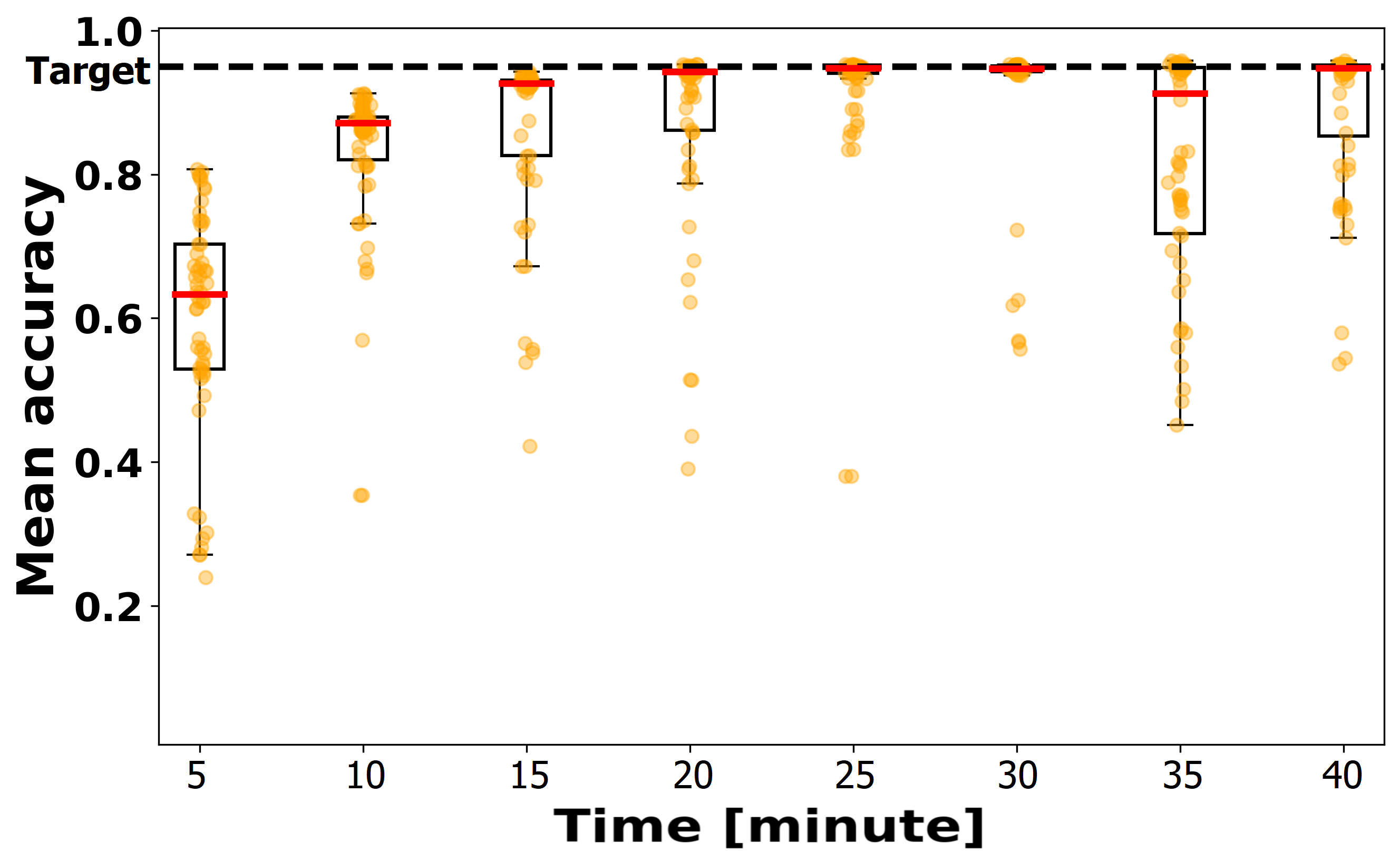}
 \caption{\small Distribution of mean accuracy at every five-minute interval for the OGL algorithms in the Luxembourg City off-peak scenario (12:00-12:40 PM) using the MNIST dataset. Each point is the mean accuracy of a single vehicle, averaged in the time interval. The target line shows the target accuracy ($95\%$)  obtained using the centralized ML training on the union of the local dataset of all vehicles. Time on the x-axis corresponds to the time from the beginning of the scheme.\normalsize}
\label{fig:box_plot_accuracy_churn}
 \end{figure}

To assess the effectiveness of our \ac{OGL} approach in dynamic settings, we consider a set of $V$ nodes which need to perform a handwritten digit recognition task (MNIST dataset~\cite{lecun_gradient-based_1998}) or object recognition (CIFAR-10 dataset~\cite{cifar}).
We assume that each node in the system is endowed with a \textit{local dataset} of different sizes and randomly selected without replacement from the original MNIST and CIFAR-10 dataset. The resulting dataset size, denoted by $d_v$, falls within the range of 50-350 samples for each node. This implied a training set size ranging from 600 kB to 3.2 MB when utilizing the CIFAR-10 dataset and between 224 KB and 645 KB when employing the MNIST dataset.
Let us denote the aggregate local dataset across all nodes in the system as the \textit{global dataset}. The size of the global dataset is 700 samples in all scenarios unless stated differently. We aim to ensure that the results remain consistent and are not subject to significant variation due to differences in global information in different scenarios.

In Problem \ref{prob:1}, the coefficient $\beta$ has been set to 1, to ensure both computing and communication costs are equally weighted. The target accuracies have been set to 0.8 and 0.4 for the MNIST and CIFAR-10 datasets, respectively. These targets were set based on the convergence accuracy of a centralized ML model trained on the global dataset. 
Furthermore, we assume the cost per byte of d2d transfer to be four times less costly than device-to-server transfers. This is because d2d transfers bypass the need for data routing or server maintenance, making them a more cost-effective solution.
We associate a \textit{global test set} obtained by random sampling $20\%$ of the source datasets and ensuring that the local datasets and test set are disjoint. We assume that nodes use a CNN model to perform both inference tasks. 

The first layer of the CNN model is a Conv2D layer, which applies a number of convolutional operations to the input image and uses the activation function  ReLU. This layer is followed by a MaxPooling2D layer with a 2x2 pool size, which reduces the spatial dimensions of the input. The Flatten layer then transforms the 2D matrix data into a 1D vector. Subsequently, a Dense layer using ReLU activation and He-uniform weight initialization is added. The final layer is another Dense layer, with the number of neurons equal to the number of output classes, and the softmax activation function is used for multi-class classification. The model is compiled with the Stochastic Gradient Descent (SGD) optimizer, and categorical cross-entropy loss function. The values of the parameters of each layer are mentioned in  Table \ref{tab:CNN_parametrs}. We choose this architecture and parameter values based on two key considerations. Firstly, some choices are widely recognized as effective in extracting shape features from images for the considered datasets~\cite{alzubaidi2021review}. Secondly, we tune some other parameters empirically by conducting a series of experiments. 

With the specified parameters, the resulting model size is approximately $320$ KB when trained on the MNIST dataset and $1.2$ MB when trained on the CIFAR-10 dataset.
Note that better energy performance might be achieved by tuning all of the CNN hyperparameters, as they impact the size of the model to be exchanged. However, this is out of the scope of the present work and is left for future developments. We end our simulations after $600$ rounds or when the average accuracy across all nodes does not improve by more than $0.5\%$ for $20$ consecutive rounds.
 
In addition to our scheme, we have considered the following baseline approaches:
\begin{itemize}
        \item \textit{Centralized ML}. In this approach, a central server possesses a dataset identical to the global dataset of the scenario, over which it trains the CNN model.
        \item \textit{Federated Averaging (Fed AVG)} \cite{mcmahan2017communication}. In this training scheme, a parameter server collects the CNN models trained locally by each node at every round, merges them, and sends the resulting CNN to each node for a new round of local training. For fairness of comparison, we assumed random client subsampling, with an average number of selected clients coinciding with the average number of nodes each node comes in contact with during a round.
    \item \textit{Decentralized Powerloss (DP)} \cite{dinani2022vehicle} 
       is a decentralized learning approach. In this approach, all the nodes set the number of local training epochs to $1$ and merge the models from all neighbours at a given time slot, without exception. In this approach, the weights associated with each model to be merged are derived from a measure of the received models' performance over the node's validation set.
      \item \textit{Random GL} is derived from \ac{OGL} algorithm, by setting uniformly at random ( and independently for each node and time slot) parameters $\mathcal{K}_{v,t}$ and $\mathcal{Z}_{v,t}$, i.e. the number of training epochs and the set of neighbour nodes whose models have to be merged.
     \item \textit{Local only}, in which each node trains the local model only on its local dataset, with no data or models exchanged with neighbouring nodes or a server.\end{itemize}
The size of local datasets varies across nodes, leading to an uneven distribution of classes. It requires using various performance metrics to compare the effectiveness of our algorithm with baseline methods.

In the first set of experiments, we considered scenarios with different numbers of nodes in the network, specifically $|V| = [3, 6, 12]$. In addition, we model the connectivity graph resulting from node mobility via an Erdős-Rényi dynamic random graph \cite{10.1145/3565287.3616530}. It is thus a sequence of graphs, each associated with a time slot. In this type of graph, an edge is established between two nodes with probability $p$, independent from other edges. Then, at each time slot, the connectivity graph stays constant, but possibly the set of edges in the graph (connection among nodes) varies.
The degree of connectivity in the network is determined by the parameter $p$. A mesh network is formed when $ p = 1$, while $p= 0.1$ leads to a sparse network.  The choice of this type of graph enables a controlled and systematic modification of node numbers, connection patterns, and node interaction frequency and duration \cite{10.1145/3565287.3616530}.

Figure \ref{fig:cost_comparisons} shows the mean total amount of computing and communication costs at convergence for OGL as well as for the baselines in different network configurations utilizing the MNIST dataset. Figure \ref{fig:cost_acc_cifar} illustrates the mean total amount of computing and communication costs at convergence for OGL as well as for the baselines using the CIFAR-10 dataset. These results suggest that our OGL scheme is by far the most energy-efficient among the gossip learning schemes, particularly concerning communication costs, achieving a level of efficiency comparable to that of Federated Learning. This confirms that context-aware tuning of the local training and merging phases of GL schemes may have a high impact on the efficiency and effectiveness of the training process.
Another key aspect resulting from our experiments is the relative mean training performance of our OGL scheme in terms of model accuracy at convergence. 
As Table \ref{tab:mnist_metrics} shows, using MNIST dataset   \ac{OGL} outperforms all baseline distributed approaches in all performance metrics, achieving performance comparable to centralized training. Critically, though being significantly more energy efficient, at convergence, OGL improves by more than $40\%$ both mean accuracy and mean loss with respect to DP, i.e. to the best performing gossip-learning approach in the state-of-the-art. Indeed, the two other distributed learning models, DP and random GL, as well as the local-only approach, fail to achieve the target mean accuracy. Figure \ref{fig:comparative_acc_cifar} depicts the mean accuracy versus round (learning process) of the \ac{OGL} and other baseline algorithms using the CIFAR-10 dataset. The outcomes derived from the CIFAR-10 dataset largely mirror those obtained from the MNIST dataset. It reinforces the effectiveness and consistency of the \ac{OGL} algorithm across different datasets. 

Figure \ref{fig:mnist_comparisons}, and  \ref{fig:cifar_comparisons} show the impact of network connectivity and the number of nodes in the system on the evolution of mean accuracy over the learning round for  MNIST and CIFAR-10 datasets.  In a system with very few nodes, the impact of optimally choosing the neighbours' contributions is relatively modest, with the mean accuracy of Random GL and DP eventually matching that of \ac{OGL}. In larger systems, our OGL tuning approach is key to achieving faster convergence and higher accuracy in sparse and dense networks. Figure \ref{fig:bar_cost_acc_cifar} illustrates that OGL maintains superior energy efficiency, notwithstanding an accuracy comparable to Random GL. Note that all schemes perform sensibly worse in the CIFAR-10 dataset, as for the same average local dataset size, its samples are more complex (i.e. larger pictures with more pixels). 

To evaluate the effectiveness of our \ac{OGL} approach in a realistic scenario,  we consider a scenario where moving nodes are vehicles traversing a region of interest.  We focused on a specific area in the city centre of Luxembourg City. This area, a square with sides measuring 1 km, was observed during a low-traffic period (off-peak) from 12:00 PM to 12:40 PM. During this time interval, there are 492 vehicles in the region, with an average sojourn time of 2.9 minutes. On average, there are about 27.3 vehicles in the region at any given time. In this scenario, vehicles are in contact if they are within each other transmission radius. The transmission radius has been set to 150 m (e.g. typical of DSRC in urban environments ~\cite{abboud2016interworking}).  In this case, on average, each vehicle is in contact with 6.7 vehicles at any time. Note that, unlike previous scenarios, the set of nodes in the region may change at different time intervals. Figure \ref{fig:histogram_churn} depicts the percentage of arrivals and departures at every 5-minute interval. In order to ensure a dynamic neighbourhood pool and give each vehicle enough time to train its local model, vehicles interchange both the loss values and the trained models at regular intervals of twenty seconds. Figure \ref{fig:box_plot_accuracy_churn} shows the mean accuracy of the vehicles approach the target accuracy, obtained by training a centralized ML model on the union of all vehicles' datasets, after ten minutes despite having churn in the network. In addition, we observe the adaptive learning capability of new arrivals. New arrivals are characterized by their initial impact on reducing the mean accuracy, as they are identified as outliers within the given time shown in Figure \ref{fig:box_plot_accuracy_churn}. Despite the initial disruption, these outliers demonstrate a capacity to learn from the existing vehicles, thereby gradually aligning with the overall trend. This is evidenced by the subsequent decrease in the number of outliers from time interval 15-20 to 20-25 minutes. This adaptive learning capability of new arrivals contributes to the robustness and resilience of the system, enabling it to maintain overall accuracy over time. It indicates our \ac{OGL}  model is able to maintain high accuracy even in the face of network instability.

%% file: conclusion.tex
\section{Conclusions}
\label{sec:conclusion}
This work presents a novel approach to an energy-efficient gossip learning scheme for dynamic settings. We employ an auxiliary  DNN model trained by an orchestrator to adaptively tune some of the key parameters of the learning process in a decentralized manner. Results indicate that our approach efficiently achieves accuracy comparable with a centralized ML method across a variety of network conditions, utilizing time-varying random graphs and a measurement-based dynamic urban scenario across two distinct datasets.\\
For future work, 
we plan to enhance the scalability and adaptability of our optimization system by developing a fully distributed optimization system, eliminating the need for an orchestrator, where nodes can self-optimize in response to drastic changes in the environment. Furthermore, we plan to investigate the impact of different types of DNN  models on the optimization and learning process, as DNN models vary in their computational requirements. This could be particularly important in a distributed learning context where computational resources are limited.